\documentclass[useAMS,usenatbib]{mn2e}
\usepackage{graphicx}
\usepackage{BibDef}
\usepackage[english]{babel}
\usepackage{natbib}
\usepackage{color}

\title[Relativistic iron K$\alpha$ line detection in the Suzaku spectra of IC 4329A]{Relativistic iron K$\alpha$ line detection in the Suzaku spectra of IC 4329A}
\author[G. Mantovani, K. Nandra, G. Ponti]{G. Mantovani\thanks{E-mail: gmantova@mpe.mpg.de}, K. Nandra, G. Ponti
\\
Max-Planck-Institut f\"{u}r extraterrestrische Physik, Giessenbachstrasse 1, D-85748 Garching bei M\"{u}nchen, Germany}
\begin{document}

\date{Accepted 2014 April 25. Received 2014 April 15; in original form 2014 March 4}

\pagerange{\pageref{firstpage}--\pageref{lastpage}} \pubyear{2014}

\maketitle

\label{firstpage}

\begin{abstract}

We present an analysis of five Suzaku observations of the bright Seyfert1 galaxy IC 4329A. The broad energy band and high signal-to-noise ratio of the data give new constraints on the iron K$\alpha$ line profile and its relationship with the Compton hump at higher energies. The Fe K bandpass is dominated by a narrow core (EW=57$_{-3}^{+3}$~eV) at 6.4 keV consistent with neutral material. Using a physically-motivated model, our analysis also reveals the presence of a broad Iron K$\alpha$ line (EW=124$_{-11}^{+11}$~eV), most likely produced in the inner part of the accretion disk and blurred by general relativistic effects. This component is not immediately evident from the individual spectra, but is clearly present in the stacked residuals of all five observations, and has high statistical significance. This highlights the difficulty in identifying broad iron lines in AGN, even in data with very high signal-to-noise ratio, as they are difficult to disentangle from the continuum. The data are consistent with the narrow and broad iron line components tracking the Compton Hump, but do not provide clear evidence that this is the case. An additional narrow Fe~{\sc xxvi} emission line at 6.94 keV is also seen, suggesting the presence of ionized material relatively distant from the central region. There is also a hint of variability, so the precise origin of this line remains unclear.

\end{abstract}


\section{Introduction}

The X-ray spectra of the Active Galactic Nuclei (AGN) are dominated by a simple power law with a photon index of $\Gamma\sim1.9$ (e.g. \citealt{nandra+94}, \citealt{Piconcelli+05}). This emission is thought to originate from inverse Comptonization of soft photons by a hot corona near the central black hole (e.g., \citealt{Shapiro+76}; \citealt{Sunyaev+80}; \citealt{Haardt+93}; \citealt{Haardt+94}). The optically thick and cold disk around the black hole (BH) reprocesses the X-rays producing the so-called Compton reflection Hump peaking near 20-30 keV (e.g., \citealt{Pounds+90}; \citealt{George+91}). Other signatures of this reflection are the absorption and emission lines produced by photoelectric absorption and fluorescence (\citealt{Matt+97}). The most prominent is the Fe K$\alpha$ emission line seen at 6.4 keV (\citealt{nandra+94}).

Observations with high spectral resolution, performed with \textit{XMM-Newton} and \textit{Chandra}, have revealed that a relatively narrow core to the Fe K$\alpha$ line is common in type 1 AGN,  with FWHM's of several thousand km s$^{-1}$  (e.g., \citealt{Yaqoob+04}; \citealt{Nandra+06}). The narrow component is thought to be produced in Compton-thick material distant from the black hole, such as the molecular torus (e.g. \citealt{Krolik+87}). The reflection can also arise in the inner parts of the accretion disk, producing a relativistically broadened component (\citealt{Tanaka+95}), modified by gravitational redshift  and relativistic Doppler effects (e.g., \citealt{Fabian+89}; \citealt{Fabian+02}). The analysis of the Iron K$\alpha$ line components is a key probe of the innermost region of the AGN (\citealt{Fabian+09}). Broad iron lines are expected, and observed, to be a widespread feature in bright AGN, with the observed fraction between $\sim30\%$ and $\sim80\%$ among nearby Seyfert galaxies (\citealt{Nandra+07}; \citealt{Calle+10}). Key outstanding issues are why some AGN apparently lack a disk line component (\citealt{Bhayani+11}) and whether or not the Fe K$\alpha$ emission is correlated with the associated hard X-ray reflection continuum, as it should if the line arises from optically thick material. 

IC 4329A ($z=0.01605$; \citealt{Willmer+91}) is one of the brightest Seyfert 1 AGN ($F\sim2\times10^{-10}$ erg s$^{-1}$ cm$^{-2}$), embedded in a nearly edge-on host galaxy. The nature of the iron K-complex in this source, and in particular the presence or otherwise of broad component to the iron line, has been controversial. Analyzing simultaneous \textit{ASCA} and \textit{RXTE} observations, \citet{Done+00} detected a moderately broadened (FWHM=43,000$\pm$11,000 km s$^{-1}$) Fe K$\alpha$ line with an equivalent width of EW=180$\pm$50 keV peaking at $\sim$6.4 keV. While the line is significantly broadened, it is not as expected from an accretion disk that extends down to the last stable orbit around a black hole. Similar results had previously been found based on \textit{ASCA} data by \citet{Nandra+97}. An \textit{XMM-Newton} observation reported by \citet{Gondoin+01} revealed a narrow core for the iron line (EW=43$\pm$1 eV) originating in mostly neutral material. \citet{Nandra+07} analyzed two XMM-Newton observations, finding one with purely narrow emission and another moderately broadened like the \textit{ASCA} case. A higher resolution view of the iron line complex in IC 4329A was provided using a 60 ks \textit{Chandra}-HETGS observation by \citet{McKernan+04}, who detected a narrow core for the 6.4 keV line together with an additional emission line near 6.9 keV. This double-peaked feature could be reproduced by several different models, included dual Gaussians, dual disk lines or a single disk line. Finally we note that \citet{Markowitz+06}, \citet{Nandra+07} and \citet{Tombesi+10} have all reported the possible presence of a blue shifted absorption line in an XMM-Newton observation of IC 4329A, which may be the signature of a very fast outflow. 

The aim of this paper is to study the nature of the iron line and reflection component in this Seyfert galaxy using data from the \textit{Suzaku} satellite. With its large effective area and broad bandpass this provides an ideal opportunity to determine the nature of the iron K$\alpha$ emission and its relationship to the Compton Hump above 10 keV. 

\section{Observations and data}

IC 4329A was observed 5 times with Suzaku in 2007 on August 1, 6, 11, 16, 20 with an exposure of $\sim$26 ks for each observation. X-ray Imaging Spectrometer (XIS; \citealt{Koyama+07}) and Hard X-ray Detector (HXD; \citealt{Kokubun+07}; \citealt{Takahashi+07}) event files were reprocessed with the calibration files available (2013-01-10 release), using the \textit{FTOOLS} package of lheasoft version 6.13, adopting the standard filtering criteria. Source spectra were extracted from a circular region of 250 arcsec centered on the source, while background spectra were extracted in an annular region ($r_{in}=320$ arcsec, $r_{out}=480$ arcsec). Response matrices and ancillary response files were produced with \textit{xisrmfgen} and \textit{xissimarfgen}. Data from the XIS0 and XIS3 were used, and the spectra were added together. We used the non-X-ray background (NXB) for the HXD/PIN data provided by the HXD team and extracted source and background spectra using the same good time intervals. The PIN spectra were corrected for dead time and the exposure of the background spectra was increased by a factor of 10, as required.

\section{Spectral Fitting}

\begin{figure}
\centering
\includegraphics[clip=true,width=0.45\textwidth,angle=270]{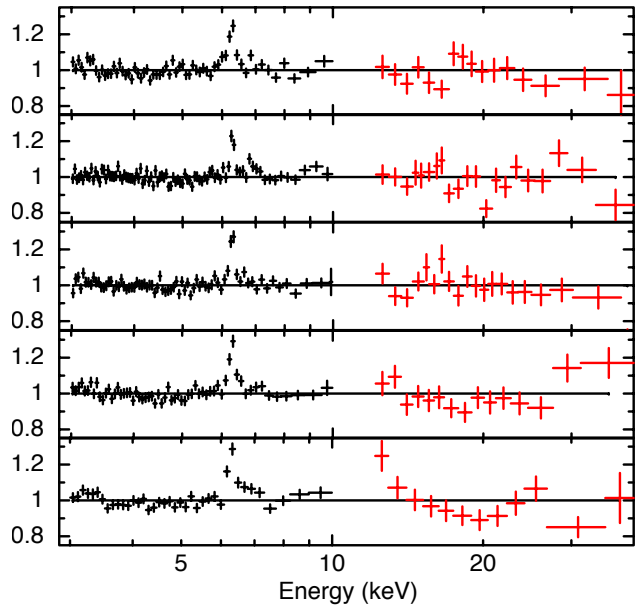}
\caption{Data/Model ratio of each observation. The model includes a neutral absorber at the redshift of the source and a reflection component. A narrow Fe K$\alpha$ line is clearly present in the data, but apparently there are no residuals consistent with a blurred component. From top to bottom, the observations are in temporal order.}
\label{power}
\end{figure}

For the spectral analysis the \textit{xspec} program (version 12.8.0) was used. All quoted errors correspond to the 90$\%$ confidence level for one interesting parameter ($\Delta\chi^2=2.71$), unless otherwise stated. 

The XIS and PIN spectra were used together in the analysis.  To minimize the effects of absorption on the spectral fits, the XIS spectra were analyzed in the 3-10 keV band, while the 12-40 keV band was chosen for the PIN spectra. We introduced a cross-normalization constant for both instruments, fixing to 0.994 the value for the XIS0 and XIS3, and to 1.164 that for the PIN, as appropriated for data taken at XIS nominal position (\citealt{Maeda+08}).

Figure \ref{power} shows the data/model ratio for each observation assuming a model with a neutral absorber at the redshift of the source (\textit{zwabs}) and a power law together with a Compton reflection component (\textit{pexrav}; \citealt{Magdziarz+95}). Residuals around 6.4 keV (Iron K$\alpha$ line) are clearly present in all the 5 observations, which we now attempt to model.

\subsection{Line models}

We tested first if this feature could be associated with emission from distant material, which produces the narrow component of the Iron K$\alpha$ line. We fitted the XIS and PIN spectra together with a model which includes a neutral absorber at the redshift of the source (\textit{zwabs}), a power law continuum together with a reflection component (\textit{pexrav}) and a Gaussian model for the iron emission line (\textit{zgauss}). The width of the Fe K$\alpha$ line was fixed to $\sigma=1$ eV. We first fitted the spectra with the high energy cut off free to vary. We measured a mean value for the cutoff of 180 keV, and subsequently adopted this as a fixed parameter in the model. The value is consistent with the recent measure of 178$_{-40}^{+74}$ keV by \citet{Brenneman+13}. 

\begin{figure}
\centering
\includegraphics[width=0.35\textwidth,angle=270]{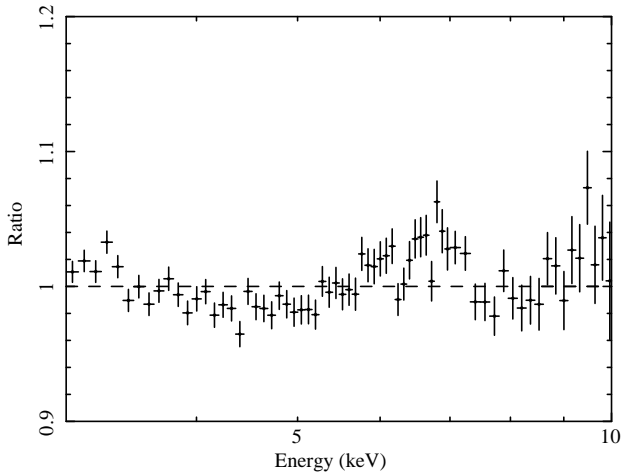}
\caption{Stacked data/model ratio for all 5 observations. Each spectrum was fitted separately with a model consisting of a single narrow Gaussian, with a neutral Compton reflection component. A broad iron line, which is not visible in the individual spectra (Fig.\ref{power}), is clearly evident in the combined residuals}
\label{1gaussXIS}
\end{figure}

\begin{figure*}
\begin{minipage}[b]{0.3\linewidth}
\centering
\includegraphics[width=1.15\textwidth]{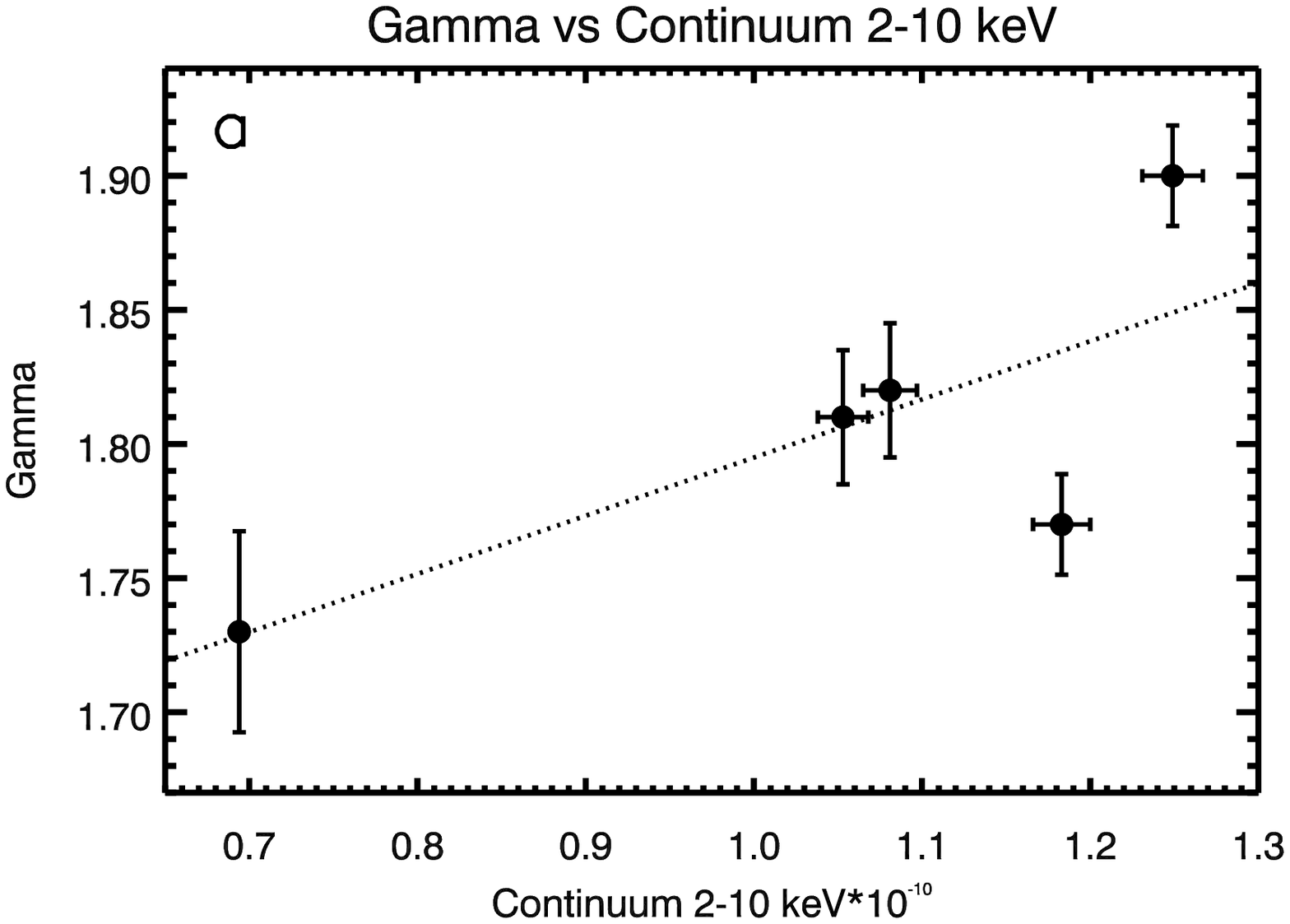}
\end{minipage}
\hspace*{0.4cm} 
\begin{minipage}[b]{0.3\linewidth}
\centering
\includegraphics[width=1.15\textwidth]{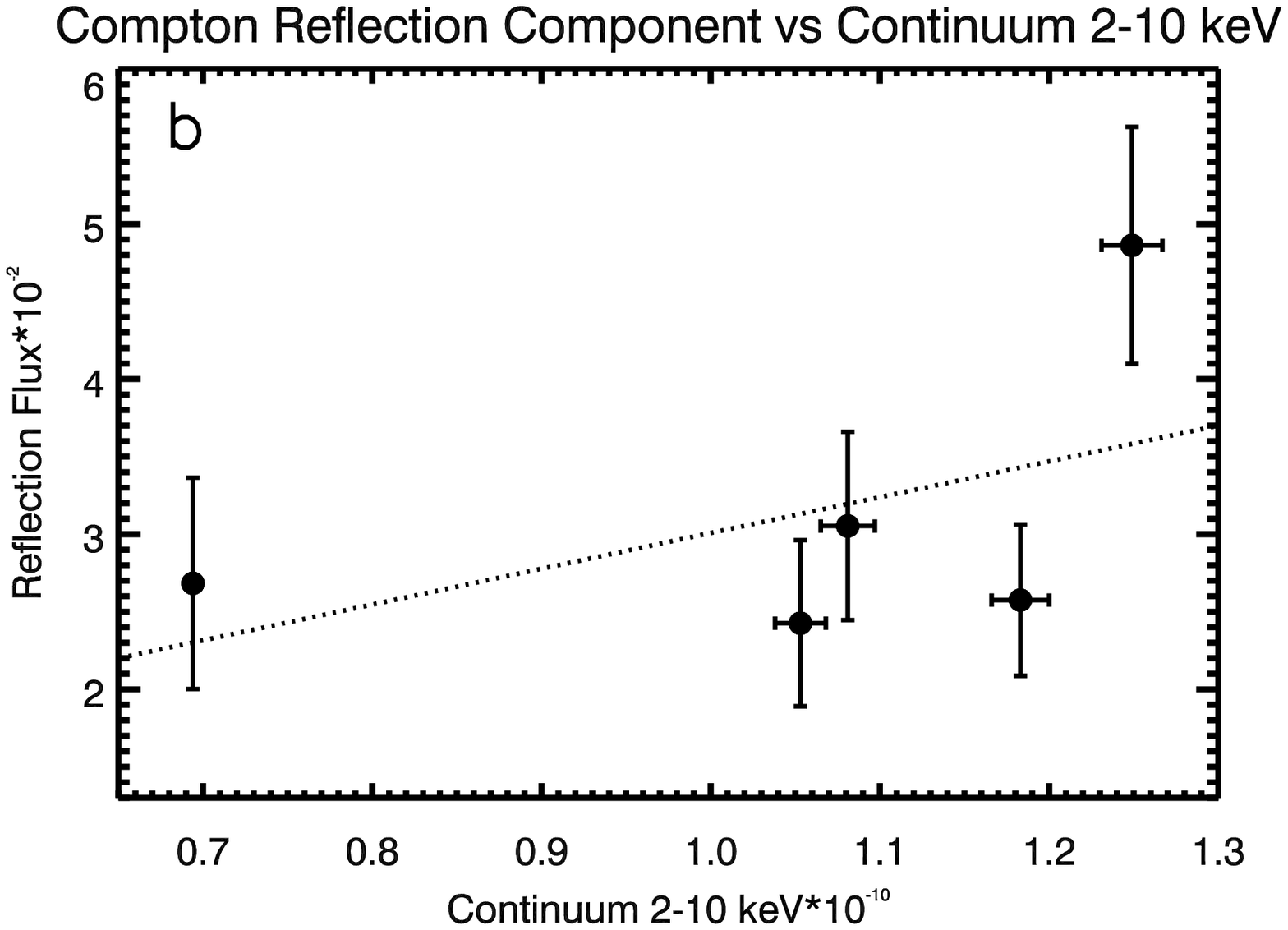}
\end{minipage}
\hspace*{0.4cm} 
\begin{minipage}[b]{0.3\linewidth}
\centering
\includegraphics[width=1.15\textwidth]{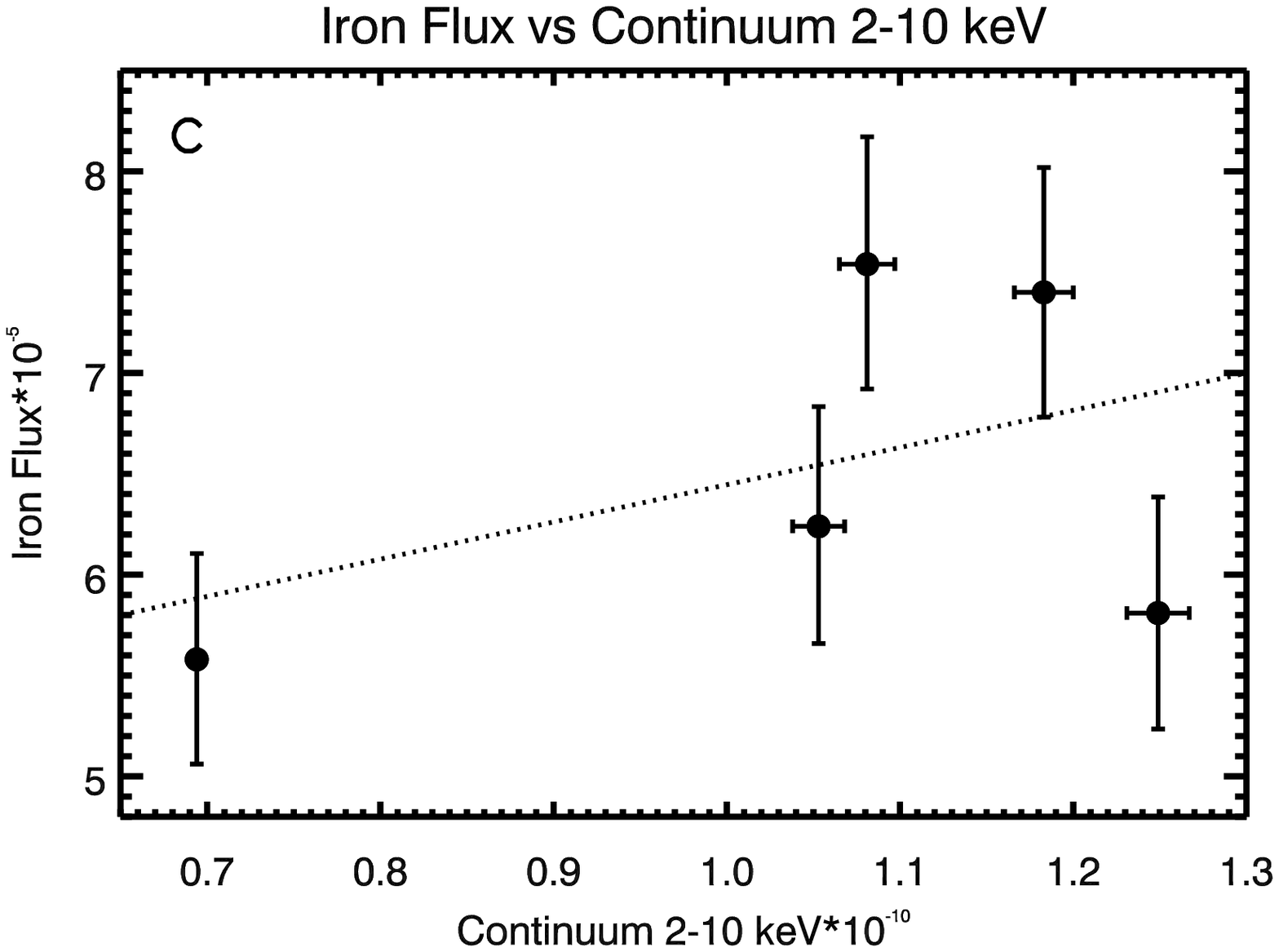}
\end{minipage}
\begin{minipage}[b]{0.3\linewidth}
\centering
\includegraphics[width=1.15\textwidth]{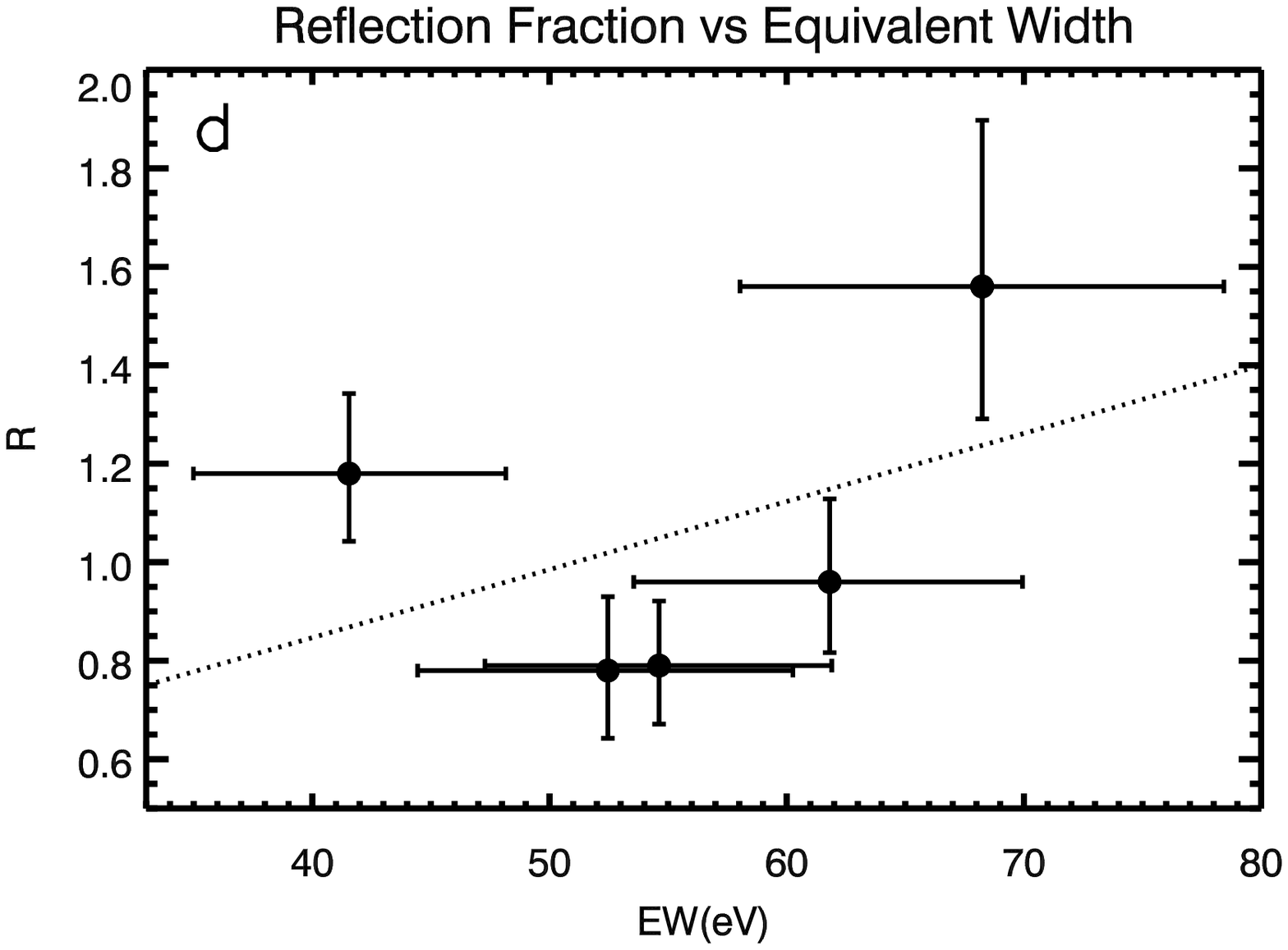}
\end{minipage}
\hspace*{0.4cm} 
\begin{minipage}[b]{0.3\linewidth}
\centering
\includegraphics[width=1.15\textwidth]{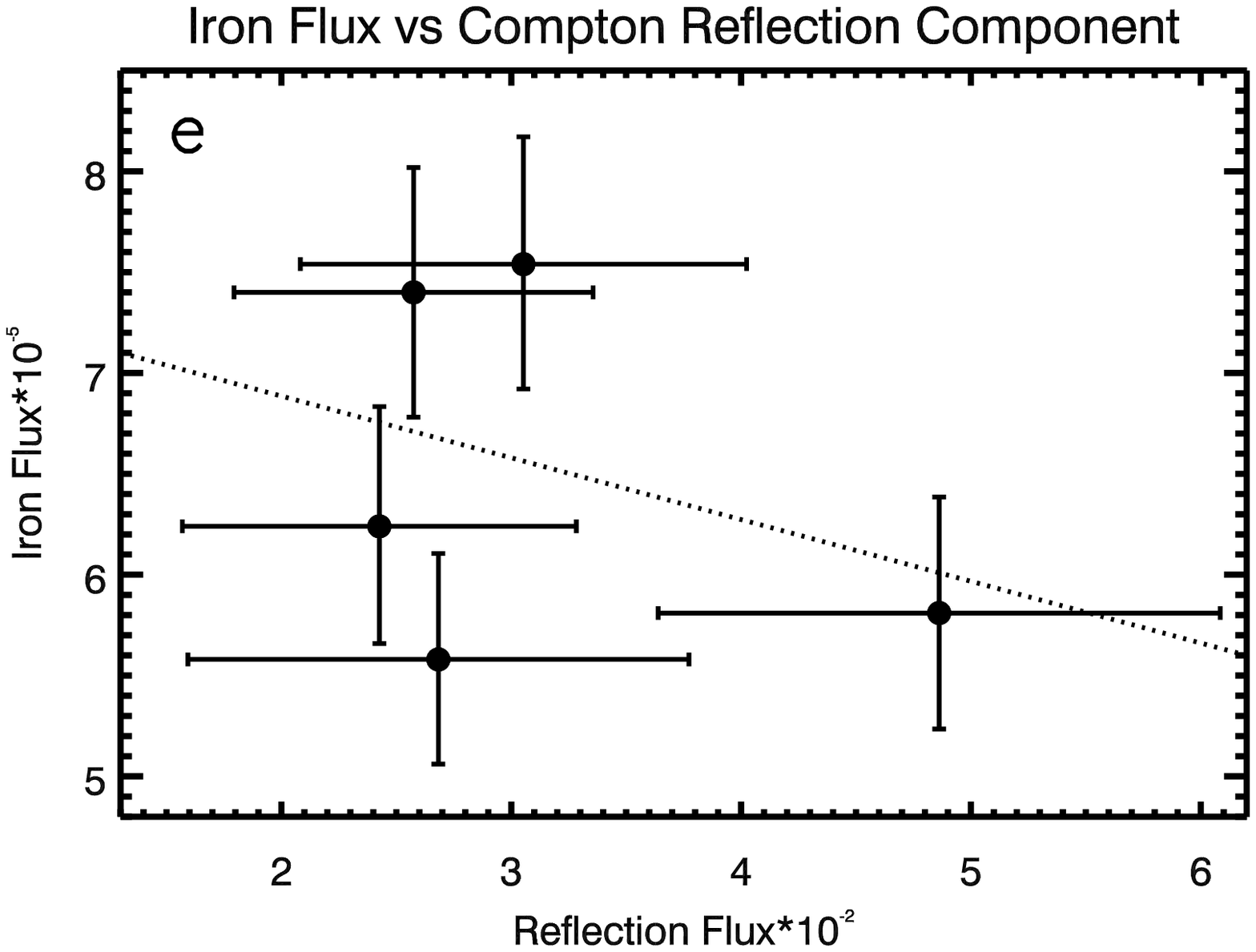}
\end{minipage}
\caption{The upper three panels (a, b, c) plot $\Gamma$, reflection flux and iron line flux as function of continuum flux in the 2-10 keV band. The bottom panels (d, e) plot the Reflection Fraction as function of Equivalent Width, the Iron flux versus Reflection Flux. The continuum fluxes are in erg cm$^{-2}$ s$^{-1}$, the Iron fluxes are in photons cm$^{-2}$ s$^{-1}$ while the Reflection Fluxes are in photons keV$^{-1}$ cm$^{-2}$ s$^{-1}$. The dotted lines show the linear fit.}
\label{corr}
\end{figure*}

Table \ref{1gaussPar} (upper panel) presents the best-fit parameters for this model (Gaussian). Figure \ref{1gaussXIS} shows the averaged data to best fit Gaussian model ratio in the 3-10 keV band. This model now reproduces well the narrow core of the Fe K$\alpha$ line, however it leaves significant residuals around 6.4 keV. In particular, a broad component extending to lower and higher energies is visible. Note that each spectrum was fitted separately, with the parameters of the continuum being free to vary in each single observation (see Table \ref{1gaussPar}). The data/model ratios from each individual fit were then combined for display purposes. This approach is therefore robust to any variations of the continuum slope, which in theory might produce artificial residuals if the spectra themselves are stacked before fitting with an average model. On the other hand \citet{Nandra+07} showed via simulations that such a variable continuum does not produce a false broad, relativistic component in any case.

The significance level of this component was estimated by introducing an accretion disk component (\textit{laor} model;  \citet{Laor+91}). The rest energy of the line was fixed to 6.4 keV, with the inner and the outer radii (R$_{in}$ and R$_{out}$) fixed to 6 and 400 $r_g$ (the inner radius being appropriate for the Schwarzschild metric) and the emissivity index fixed to $q=3$. A fixed inclination of 35$^{\circ}$ was also adopted (see below) leaving the line flux as the only additional free parameter. The improvement in $\chi^{2}$ for each individual spectrum ranged from 3-18, corresponding to 2-4$\sigma$ based on the F-test. When all the improvements $\chi^{2}$ are considered together, however, the significance level reaches $\sim$5.5$\sigma$ based on this test.

\begin{table}
\setlength{\tabcolsep}{10pt}
\resizebox{8.5cm}{3cm} {
{ \huge
\begin{tabular}{@{}llccccc@{}} \hline
&Gauss&702113010&702113020&702113030&702113040&702113050\\\hline
&$\Gamma$&$1.81\pm0.04$&$1.90\pm0.03$&$1.77\pm0.03$&$1.82\pm0.04$&$1.73\pm0.06$\\
&R&$0.78\pm0.23$&$1.18\pm0.24$&$0.79\pm0.20$&$0.96\pm0.25$&$1.56\pm0.48$\\ 
&N&$3.11\pm0.14$&$4.12\pm0.16$&$3.26\pm0.13$&$3.18\pm0.14$&$1.72\pm0.12$\\
&$E_{{\footnotesize K\alpha}}$&$6.39\pm0.02$&$6.40\pm0.01$&$6.38\pm0.02$&$6.40\pm0.02$&$6.40\pm0.02$\\ 
&F$_{{\footnotesize K\alpha}}$&$6.24\pm0.94$&$5.81\pm0.92$&$7.40\pm0.99$&$7.54\pm1.00$&$5.58\pm0.83$\\
&F$_{{\footnotesize a}}$&$1.05\pm0.02$&$1.25\pm0.02$&$1.18\pm0.02$&$1.08\pm0.02$&$0.69\pm0.01$\\ 
&F$_{{\footnotesize b}}$&$1.56\pm0.01$&$1.82\pm0.01$&$1.89\pm0.01$&$1.68\pm0.01$&$1.45\pm0.01$\\ 
&F$_{{\footnotesize CH}}$&$2.43\pm0.86$&$4.86\pm1.22$&$2.58\pm0.78$&$3.05\pm0.97$&$2.68\pm1.09$\\
&EW&$52\pm8$&$42\pm7$&$55\pm7$&$62\pm8$&$68\pm10$\\ 
&$\chi^2$&959.81/891&1117.19/1104&993.38/999&921.63/869&626.51/606\\ \hline 
&Pex&702113010&702113020&702113030&702113040&702113050\\ \hline
&$\Gamma$&$1.82\pm0.03$&$1.85\pm0.02$&$1.78\pm0.03$&$1.83\pm0.03$&$1.71\pm0.04$\\
&N&$3.15\pm0.12$&$3.97\pm0.12$&$3.31\pm0.11$&$3.25\pm0.12$&$1.69\pm0.08$\\ 
&$R_{{\footnotesize N}}$&$0.38\pm0.07$&$0.32\pm0.06$&$0.36\pm0.06$&$0.44\pm0.07$&$0.46\pm0.09$\\
&$R_{{\footnotesize B}}$&$0.29\pm0.13$&$0.23\pm0.11$&$0.26\pm0.12$&$0.36\pm0.14$&$0.47\pm0.18$\\ 
&F$_{{\footnotesize a}}$&$1.05\pm0.02$&$1.25\pm0.02$&$1.18\pm0.02$&$1.08\pm0.02$&$0.69\pm0.01$\\ 
&F$_{{\footnotesize b}}$&$1.53\pm0.01$&$1.78\pm0.01$&$1.87\pm0.01$&$1.65\pm0.01$&$1.42\pm0.01$\\ 
&$\chi^2$&942.25/891&1124.08/1104&987.78/999&899.82/869&610.57/606\\
\hline
\end{tabular}
}}
\caption{Upper panel: best-fit parameters for the Gaussian model. Lower panel: best-fit parameters for Reflection model. The continuum fluxes in the 2-10 keV (F$_{a}$) and 12-40 keV (F$_{b}$) bands are in 10$^{-10}$ erg s$^{-1}$ cm$^{-2}$ while the Normalizations (N) and the Compton Hump fluxes (F$_{CH}$) are in 10$^{-2}$ photons keV$^{-1}$ cm$^{-2}$ s$^{-1}$. The Iron K$\alpha$ fluxes (F$_{K\alpha}$) are in 10$^{-5}$ erg s$^{-1}$ cm$^{-2}$. The energy of the line (E$_{K\alpha}$) is in keV and the equivalent width (EW) is in eV. $R_{N}$ and $R_{B}$ correspond to the reflection fraction of the narrow and broad Iron line component, respectively.}
\label{1gaussPar}
\end{table}

\subsection{Reflection model}

To provide a more physically self-consistent model for the spectra, we use the \textit{pexmon} model (\citealt{Nandra+07}), which accounts for the Compton reflection and iron K$\alpha$ emission from neutral, optically thick material in a slab geometry. This model combines: i) narrow Fe K$\alpha$ at 6.4 keV; ii) narrow Fe K$\beta$ at 7.06 keV; iii) Ni K$\alpha$ at 7.47 keV; iv) Compton Reflection (as the \textit{pexrav} model) and v) Fe K$\alpha$ Compton shoulder. The Fe K$\beta$ and Ni K$\alpha$ line fluxes are 11.3$\%$ and 5$\%$ respectively of that for Fe K$\alpha$ (\citealt{George+91}). 
 
Together with a neutral absorber at the redshift of the source (\textit{zwabs}) and a cut-off power law (\textit{cutoffpl}) which models the continuum, we included two \textit{pexmon} components, in order to account for the narrow core of the line, and to test if any broadened component is present. For the former, we fixed the inclination of the slab to 60$^{\circ}$, parametrizing the strength of the reflection with the $R$ parameter, the ratio of the reflection normalization to the power law component.  For the latter, the \textit{pexmon} model is convolved with the \textit{kdblur2} model (\citealt{Crummy+06}), which broadens the reflection component appropriately to account for relativistic effects close to the black hole. Following \citet{Nandra+07} we parametrized the emissivity of the reflection by a broken power law with fixed indices of $q=0$ and $q=3$ above and below a break radius $R=R_{break}$. The R$_{break}$ value was fixed to 20. The inner and the outer radii (R$_{in}$ and R$_{out}$) were fixed to 6 and 400 $r_g$, as above. Abundances were fixed to solar in both reflection models, and the spectral index $\Gamma$ is tied to that of the power law. We first fitted the spectra with the inclination of the blurred component free to vary, but subsequently fixed this parameter to 35$^{\circ}$, the mean value for all 5 observations.  

In Table \ref{1gaussPar} (lower panel) we report the best fit parameters for this model. The spectral indices ($\Gamma$) are not consistent within the errors between the observations, indicating spectral variation of the continuum. It can also be seen that both the narrow and broad reflection components are clearly present in each observation, as indicated by the fact that their strengths ($R_{N}$ and $R_{B}$) are well constrained in the fits. This confirms the conclusion from the visual inspection of Fig.~\ref{1gaussXIS} that a relativistically blurred component is present in these data. 

In addition to the narrow 6.4 keV core and the broad component, two of the spectra (702113020 and 702113050) show strong evidence for a Fe~{\sc xxvi} narrow emission line in the observations. Adding this to the model with the two reflection components yields line energies of $6.94_{-0.01}^{+0.04}$ and $6.94_{-0.13}^{+0.04}$, respectively, consistent with a narrow Fe~{\sc xxvi} line emitted from distant material. Including this line in the model, Table \ref{chiFe} shows the $\chi^2$ of the narrow Gaussian model and the more physical reflection model, and the $\Delta\chi^2$ between them. Generally the latter gives a much better fit to the spectra (the exception being 702113020, where the $\chi^{2}$ is similar). 

Finally, we tested for the presence of a high energy absorption line around 7.5~keV, as first reported by \citet{Markowitz+06}. None of the {\it Suzaku} spectra shows any evidence for such a feature. 

\begin{table}
\centering
\begin{tabular}{cclcclcclcc|l}\hline
&Observation& Gaussian & Reflection &$\Delta\chi^2$\\ \hline
&702113010&950.20/890&933.88/890&16.32\\ 
&702113020&1097.46/1103&1100.35/1103&-2.89\\ 
&702113030&993.35/998&987.78/998&5.57\\ 
&702113040&921.33/868&899.79/868&21.54\\ 
&702113050&616.26/605&600.41/605&15.85\\ 
\end{tabular}
\caption{Comparison of the $\chi^2$ of Gaussian and Reflection model when the highly ionised emission line component at 6.94 keV is also included.}
\label{chiFe}
\end{table}

\subsection{Spectral variability}

The energy of the Gaussian line (E$_{K\alpha}$) is always consistent, within the errors, with the neutral Fe K$\alpha$ transition (Table \ref{1gaussPar}, upper panel). The continuum flux (F$_{a}$), power law slope ($\Gamma$), reflection fraction $R$  and iron K$\alpha$ flux (F$_{K\alpha}$) all show evidence for variability between the different observations.  A correlation between the slope of the power law and the continuum flux in the 2-10 keV band is commonly observed in AGN spectra (\citealt{Leighly+96}; \citealt{Lamer+00}; \citeyear{Lamer+03}; \citealt{Ponti+06}). Some degree of correlation is also seen here (Figure \ref{corr}a), but the significance is low ($\sim70$\% confidence). Figure \ref{corr}b and Figure \ref{corr}c show the relationship between the reflection flux (F$_{CH}$) and the Fe K$\alpha$ flux (F$_{K\alpha}$) with the continuum (F$_{a}$), respectively. Contrary to expectations, no clear correlation between these parameters is observed. Figure \ref{corr}d and \ref{corr}e also shows no clear correlation is present also between the reflection fraction (R) and either the equivalent width (EW), or the Fe K$\alpha$ line flux (F$_{K\alpha}$).

\section{Discussion}

We have analyzed 5 Suzaku observations in order to determine the nature of the Fe K$\alpha$ emission and its relation to the Compton Hump in the Seyfert 1 galaxy IC 4329A. The analysis of the iron line shows a core with an energy that is consistent with an origin in neutral reflection form distant material. Once this narrow component of the line is fitted, broad residuals in the Fe K band appear, but these are only clearly evident  when all datasets are combined (Fig. \ref{1gaussXIS}): they are not clearly visible in the single data/model ratio of each observation (Fig. \ref{power}). These broad residuals are well reproduced by a model appropriate to an origin in the inner accretion disk, where the line and the continuum are blurred by relativistic effects. This is the first clear demonstration of a relativistic reflection component in this source, with previous reports suggesting either a narrow line only, or moderate broadening, the latter having been interpreted as evidence for a truncated accretion disk. 

The apparent lack of relativistic signatures are in the X-ray spectra of some nearby AGN -- including IC 4329A -- is a major outstanding issue for the standard accretion disc paradigm for feeding of AGN. If a relatively cool, optically thick accretion disk is present around the black hole and X-ray emission impinges on the disk, this component is hard to avoid. Disk truncation to large radii seems implausible because of the very high radiative efficiency of these systems, and the fact that most of the energy dissipation is expected to occur in the central regions. \citet{Bhayani+11} have suggested in general, and specifically for this source, that strong relativistic effects, disk ionization and/or high disk inclination can explain the apparent lack of relativistic signatures due to the difficulty of disentangling very broad features from the continuum. While these effects may be operating in IC 4329A the spectra are well fit by a relatively simple, normal relativistic, reflection component from a neutral disk in a Schwarzschild geometry seen at a relatively low inclination  ($35\deg$). The difficulty in recognizing the relativistic component in IC 4329A stems from its relative weakness compared to the expectations for a semi-infinite flat disk geometry illuminated by a point source ($R_{B}\sim 0.3$ compared to $R\sim 1$). It has been noted, however, that the latter geometry is that which produces the very strongest reflection (e.g. \citealt{Murphy+09}), and any significant deviation from this results in lower reflection fractions. General relativistic effects close to the black hole can also result in reflection that is either stronger, or weaker, than this expectation (\citealt{Miniutti+03}). Either or both effects could be at play in IC 4329A. 

Our work supports the contention of \citet{Calle+10} and \citet{Nandra+07} that very high signal-to-noise ratio is required to disentangle broad iron line components even in nearby AGN. Despite IC 4329A being the second brightest Seyfert galaxy in the 2-10 keV bands, it was necessary to sum all the data/model ratios and reach a total exposure time of $\sim$130 ks in order to clearly detect the important residual at the energy of the Iron line. 

Another aspect of our current work is the examination of the relationship between the iron line strength and the Compton Hump. Being features of the same reflected spectrum, they should vary together from spectrum-to-spectrum. Fits with the phenomenological Gaussian line model have revealed no clear relationship between these components in our fits. On the other hand, the physically motivated reflection model, in which the line is forced to follow the reflection continuum, provides a much better fit to the data, suggesting that such a correlation might in fact hold. As a further test, we allowed the iron abundance to be free in the reflection model. While largely unphysical (one would not expect the iron abundance to vary on these short timescales), the effect of this is to decouple the emission line from the reflection continuum, while retaining the requirement for both to be present, and accounting for the fact that there is evidence in the spectra for both a blurred and distant reflection component. Applying such a model we find consistency in the iron abundance for all spectra, with tentative evidence that it is sub-solar ($A_{Fe}\sim 0.5$). If this is the case, a low iron abundance could be another factor contributing to the difficulty of detecting the broad emission line  in this source. Regardless, the consistency of the iron abundance between the five observations shows that the spectra are consistent with the iron line tracking the reflection continuum, as expected in the standard model. 

We also detected a narrow emission at 6.94 keV the 99.7$\%$ significance in two observations, in agreement with the work of \citet{McKernan+04}. This feature does not appear to be the blue wing of a disk line, so is more likely to be associated with highly ionized gas, and identified with the Fe~{\sc xxvi} emission line, albeit with a small redshift.

\section*{Acknowledgments}

We thank the anonymous referee for constrictive comments on this manuscript. This research has made use of data obtained from the Suzaku satellite, a collaborative mission between the space agencies of Japan (JAXA) and the USA (NASA). G.P. acknowledges support via an EU Marie Curie Intra-European Fellowship under contract: FP7-PEOPLE-2009-IEF-254279.

\bsp

\label{lastpage}

\end{document}